\documentstyle[tighten,floats,aps,epsf]{revtex}

\begin{document}
\title
{Deformation of C isotopes}

\author{Y. Kanada-En'yo}

\address{Institute of Particle and Nuclear Studies, \\
High Energy Accelerator Research Organization,\\
Ibaraki 305-0801, Japan}

\maketitle
\begin{abstract}
Systematic analysis of the deformations of proton and neutron densities in 
even-even C isotopes was done based on the method of antisymmetrized 
molecular dynamics. 
The $E2$ transition strength was discussed 
in relation to the deformation.
We analyze the $B(E2;2^+_1\rightarrow 
0^+_1)$ in $^{16}$C, which has been recently measured to be
abnormally small.
The results suggest the difference of the deformations 
between proton and neutron densities in 
the neutron-rich C isotopes.
It was found that stable proton structure in C isotopes plays an 
important role in the enhancement the neutron skin structure as well as
in the systematics of $B(E2)$ in the neutron-rich C. 
\end{abstract}

\noindent

\section{Introduction}
In light unstable nuclei, exotic phenomena such as 
neutron halo and neutron skin structures were discovered
owing to the progress of the experimental technique.
These contradict to the traditional understanding for 
stable nuclei where the proton density and neutron densities are
consistent with each other in a nucleus. These phenomena imply that
the exotic features may appear in unstable nuclei
due to the difference between proton and neutron densities.
Another subject concerning the difference is 
the deformations of proton and neutron densities. 
For example, the opposite deformations between proton and neutron densities in 
proton-rich C isotopes were theoretically suggested \cite{ENYO-c10}. 

Recently, the life time of the $2^+_1$ state of $^{16}$C 
has been measured\cite{Imai04}.
It indicates the abnormally small $E2$ transition strength as
$B(E2;2^+_1\rightarrow 0^+_1)=$0.63 e$^2$fm$^4$ in $^{16}$C, 
compared with those for other C isotopes($^{10}$C, $^{12}$C 
and $^{14}$C).
As well known, $B(E2)$ is related to the intrinsic deformation of the 
nucleus. Considering the excitation energy $E_x(2^+_1)=1.766$ MeV of $^{16}$C, 
it is expected that this nucleus is not spherical but has a deformed structure.
In case of normal stable nuclei, $B(E2)$ values are generally large 
in the deformed nuclei. It means that the hindrance of the 
$B(E2;2^+_1\rightarrow 0^+_1)$ in $^{16}$C seems to contradict to the 
deformation expected from the $E_x(2^+_1)$.
Moreover, the neutron and proton transition matrix elements, $M_n$, $M_p$ 
derived from the $^{208}$Pb+$^{16}$C inelastic scattering\cite{Elekes04} 
imply that the neutron excitation is dominant in the $2^+_1$ state
of $^{16}$C.
From the point of view of collective deformations, these experimental 
results suggest the possible 
difference between proton and neutron shapes in $^{16}$C.

In the theoretical work on the proton-rich C isotopes\cite{ENYO-c10}, 
it was suggested that the difference of the shapes between proton 
and neutron densities may cause the suppression of $B(E2)$.
In the previous work \cite{ENYO-c10},
the difference between the proton and neutron shapes 
in the neutron-rich C isotopes was predicted as well as the proton-rich side.
It is natural to consider that 
the systematic analysis of the proton and neutron deformations 
in connection to the $B(E2)$ 
is important to understand the properties of 
the neutron-rich C. It may be helpful also to predict 
exotic phenomena in unstable nuclei.

In this paper, we study the deformations of proton and
neutron densities in C isotopes based on the 
theoretical calculations with antisymmetrized molecular dynamics(AMD).
The AMD method is a useful approach for the structure study 
of stable and unstable nuclei. The applicability of this method in 
the systematic analysis of the light nuclei has been proved in many
works \cite{ENYObc,ENYOsup,AMDrev,Itagaki00,Thiamova03}.
Especially, in the description of the deformation and clustering aspect
in light nuclei, this method has an advantage over the 
normal mean-field approaches like Hartree-Fock calculations. 
which sometimes fail to describe
the deformation in the very light nuclei such as $^{12}$C.
Although shell model calculations are useful 
to investigate the level structure of light nuclei,
they are not suitable for the systematic study of the deformations 
because of the difficulty in directly extracting intrinsic deformations
as well as the ambiguity of the effective charges in the unstable nuclei. 
In order to extract a naive picture on the proton and neutron shapes
we apply the simplest version of AMD, based on a single AMD wave function. 
By using the AMD method, 
we analyze the systematics of the deformation and $E2$ transition strength
in neutron-rich C isotopes. The hindrance of the 
$E2$ transition strength in $^{16}$C is discussed.
The theoretical predictions for the deformation and $B(E2)$ in 
further neutron-rich isotopes, $^{18}$C and $^{20}$C, are also reported.

This paper is organized as follows. 
In section \ref{sec:formulation}, the formulation of AMD is 
briefly explained. We show the results of the energies, radii, and $B(E2)$ 
obtained by the simple version of AMD,
while comparing them with the experimental data in Sec.\ref{sec:results}. 
In Sec.\ref{sec:discuss}, the intrinsic deformations of 
proton and neutron densities are analyzed in connection to the 
observable such as the $E2$ and radii.
In Sec.\ref{sec:vap}, 
we show the results of $^{16}$C obtained by an extended version of AMD.
Finally, a summary is given in Sec.\ref{sec:summary}.

\section{Formulation}
 \label{sec:formulation}

The formulation of AMD 
for nuclear structure studies is explained in \cite{ENYObc,AMDrev,ENYOe}.

The wave function of a system with the mass number $A$ 
is written by a superposition of AMD
wave functions $\Phi_{AMD}$. An AMD wave function is given by a single
Slator determinant of Gaussian wave packets as,
\begin{equation}
 \Phi_{\rm AMD}({\bf Z}) = \frac{1}{\sqrt{A!}} {\cal{A}} \{
  \varphi_1,\varphi_2,...,\varphi_A \},
\end{equation}
where the $i$-th single-particle wave function is written as,
\begin{eqnarray}
 \varphi_i&=& \phi_{{\bf X}_i}\chi_i\tau_i,\\
 \phi_{{\bf X}_i}({\bf r}_j) &\propto& 
\exp\bigl\{-\nu({\bf r}_j-\frac{{\bf X}_i}{\sqrt{\nu}})^2\bigr\},
\label{eq:spatial}\\
 \chi_i &=& (\frac{1}{2}+\xi_i)\chi_{\uparrow}
 + (\frac{1}{2}-\xi_i)\chi_{\downarrow}.
\end{eqnarray}
In the AMD wave function, the spatial part is represented by 
complex variational parameters, ${\rm X}_{1i}$, ${\rm X}_{2i}$, 
${\rm X}_{3i}$, which indicate the center of the Gaussian wave packets.
The orientation of the intrinsic spin is expressed by
a variational complex parameter $\xi_{i}$, and the iso-spin
function is fixed to be up(proton) or down(neutron).

In order to obtain a naive understanding on the systematics of
intrinsic deformations, we use a simplest version of 
the AMD method which was applied 
for Li, Be and B isotopes in Ref.\cite{ENYObc}.
Namely, we perform energy variation for a 
parity-eigen state, $P^\pm\Phi_{AMD}\equiv \Phi^\pm_{AMD}$,
 projected from an AMD wave function. 
We consider the AMD wave function obtained by the energy variation 
as the intrinsic state, and 
the total-angular-momentum projection($P^J_{MK}$) is 
done after the variation to evaluate observables 
such as the energies, radii, and the transition strength. 
Thus the variation is done after the parity projection, but the 
total-angular-momentum projection is performed after the variation. 
This method is called as 
VBP(variation before projection) in the present paper. 
For further investigations of the level scheme of $^{16}$C, 
we also perform the VAP(variation after projection) calculation
with respect to both the parity and total-angular-momentum projection
as the same way as done in Refs.\cite{ENYOe,ENYOg}.
In the VBP calculations, we fix the orientation of the intrinsic spin 
$\xi_{i}$ to be up or down. In the VAP calculations, $\xi_{i}$'s are treated as
free variational parameters.

\section{Interactions} 
\label{sec:interaction}

The effective nuclear interactions adopted in the present work
consist of the central force, the
 spin-orbit force and the Coulomb force.
We adopt MV1 force \cite{TOHSAKI} as the central force.
This central force contains a zero-range three-body force
as a density-dependent term in addition to the two-body interaction.
The Bertlett and Heisenberg terms are chosen to be $b=h=0$.
We use the parameter set, case 3 of MV1 force with the Majorana 
parameter as $m=0.576$, which was adopted 
in Ref.\cite{ENYObc}.  
Concerning the spin-orbit force, the same form of the two-range Gaussian 
as the G3RS force \cite{LS}
is adopted. The strengths of the spin-orbit force, 
(a) $u_{I}=-u_{II}\equiv u_{ls}=900$, and (b) 1500 MeV are used. 
In the VAP calculations, we also use the interaction parameters, 
(c) the case 1 of MV1 force with $m=0.62$ and the spin-orbit force with
 $u_{ls}=3000$ MeV used in Ref.\cite{ENYOe}.

\section{Results of VBP calculations}\label{sec:results}

The structure of positive parity states of even-even C isotopes 
are studied by the VBP calculations
within the framework of AMD.
In this section, we present theoretical results, such as the energies,
radii, and $E2$ transitions, while comparing them with the experimental data.
The optimum width parameter $\nu$ in Eq.\ref{eq:spatial} 
is chosen to minimize the energy of the system for each nucleus.  
The adopted $\nu$ parameters are listed in Table.\ref{tab:nu}.
After the variation, we perform the total-angular-momentum projection 
$P^J_{MK}$ and diagonalize the Hamiltonian and norm matrices,
$\langle P^J_{MK'}\Phi^\pm_{AMD}|H| P^J_{MK''}\Phi^\pm_{AMD}\rangle$
and $\langle P^J_{MK'}\Phi^\pm_{AMD}| P^J_{MK''}\Phi^\pm_{AMD}\rangle$
with respect to the $K$-quantum ($K',K''$).
Consequencely, we obtain $0^+_1$ and $2^+_1$ states 
which can be regarded to belong to the ground
$K^\pi=0^+$ band in each C isotope. 
In cases of $^{10}$C, $^{16}$C and $^{18}$C,
the second $2^+$ state in the $K^\pi=2^+$ band arises 
because of the axial asymmetry. 

\begin{table}
\caption{ \label{tab:nu} The adopted 
width parameter $\nu$ of the AMD wave functions for C isotopes.}
\begin{center}
\begin{tabular}{cccccccc}
 & $^{10}$C & $^{12}$C & $^{14}$C & $^{16}$C & $^{18}$C & $^{20}$C    \\ 
$\nu$ & 0.185 & 0.190 & 0.180 & 0.175  & 0.170 & 0.165 \\
\end{tabular}
\end{center}
\end{table}

\begin{figure}
\noindent
\epsfxsize=0.49\textwidth
\centerline{\epsffile{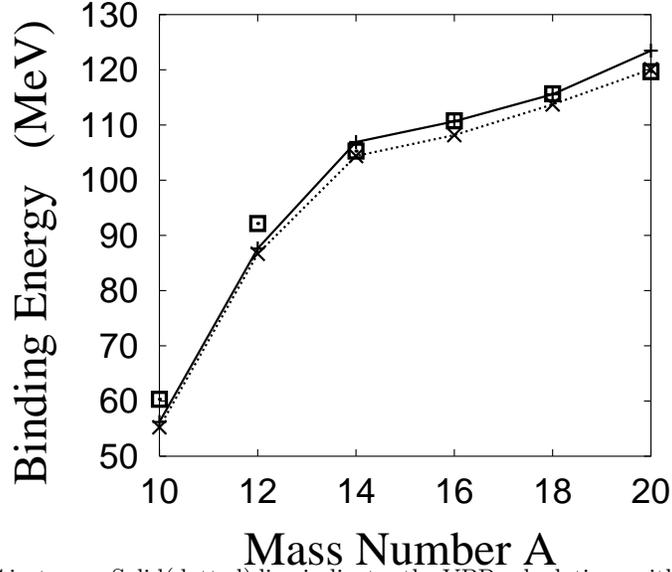}}
\caption{\label{fig:cbe}
Binding energies of C isotopes. Solid(dotted) line indicates 
the VBP calculations with $m=0.576$ and $u_{ls}=1500(900)$ MeV. 
}
\end{figure}

The binding energies and the excitation energies of the 
low-lying positive-parity states are shown in Fig.\ref{fig:cbe} 
and in Table.\ref{tab:ene}.
The binding energies of C isotopes are systematically reproduced by
the present calculations (Fig.\ref{fig:cbe}).
As mentioned before, 
the second $2^+$ state, which belong to the side band $K^\pi=2^+$,  
is found in addition to the $2^+_1$ state, 
in $^{10}$C, $^{16}$C and $^{18}$C.

The excitation energies $E_x$ of $2^+_1$ states 
tend to be underestimated by the calculations, especially in the 
neutron-rich C isotopes. This is a general tendency of the VBP calculations 
where the $0^+_1$ and $2^+_1$ states are obtained by the 
total-angular-momentum projection from a single intrinsic wave function.
It is considered to be because the wave function obtained in the VBP
may be optimized for the $2^+_1$ state while
it is relatively not enough for the $0^+_1$ state to reproduce the
level spacing between the $2^+_1$ and $0^+_1$ states.
In other words, 
the energy of $0^+_1$ state is considered to be relatively overestimated
in the VBP.
The quantitative reproduction of $E_x(2^+_1)$ is improved by VAP calculations,
where the wave function is optimized for each spin-parity state.
Based on the VBP and VAP calculations, 
the level structure of $^{16}$C is discussed later in Sec.\ref{sec:vap}.
We give a comment that also the AMD+GCM(generator coordinate method)
calculations,which is an extended version of AMD, 
well reproduce the $E_x(2^+_1)$ in neutron-rich C.

\begin{figure}
\noindent
\epsfxsize=0.49\textwidth
\centerline{\epsffile{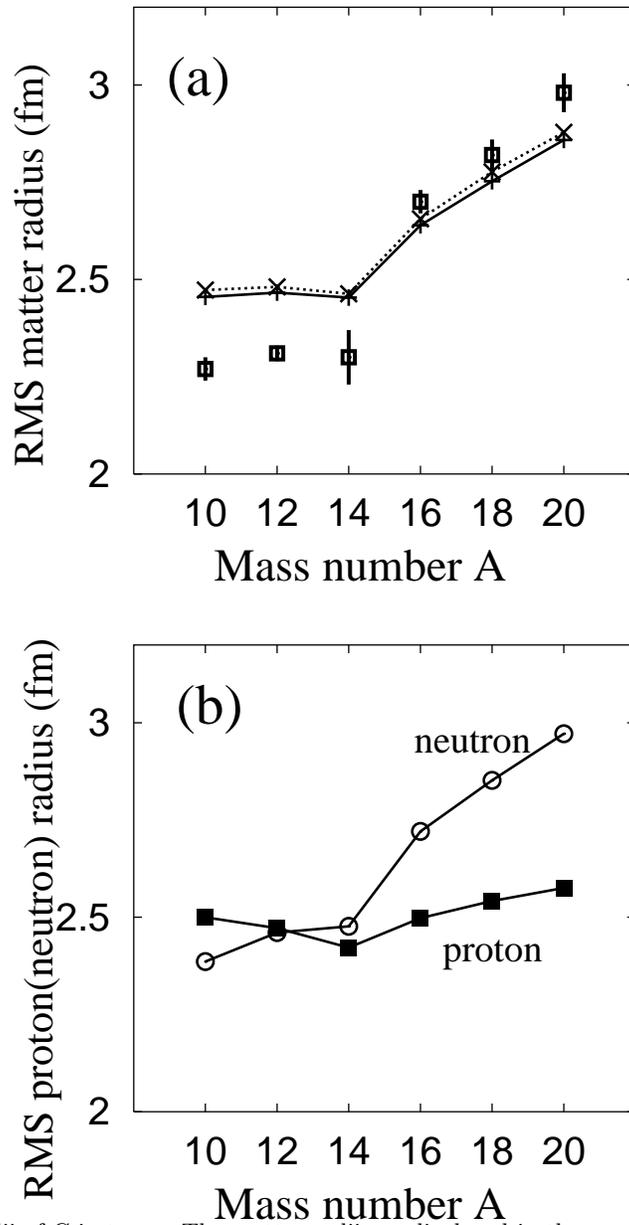}}
\caption{\label{fig:crmsr}
Root-mean-square radii of C isotopes. 
The matter radii are displayed in the upper panel(a).
The solid(dotted) line 
indicates the VBP calculations with $m=0.576$ and $u_{ls}=1500(900)$ MeV. 
The experimental data, which are derived from the interaction 
cross sections \protect\cite{Ozawa}, are shown by cross points.
The point-like proton and neutron radii calculated with 
with $m=0.576$ and $u_{ls}=1500$ MeV are shown in 
the lower panel(b).}
\end{figure}

Figure \ref{fig:crmsr} shows the results of the 
root-mean-square radii comparing with the experimental radii
derived from the interaction cross sections. 
In the systematics of the matter radii in the even-even C isotopes, there is 
a gap between $^{14}$C and $^{16}$C. This behavior is reproduced by the 
present calculations and can be described by the large deformation 
in $^{16}$C and the spherical shape due to the shell closure in $^{14}$C.
The details are discussed in the next section.

\begin{table}
\caption{ \label{tab:ene} Excitation energies of $2^+$ states of C isotopes
obtained by the VBP calculations.}
\begin{center}
\begin{tabular}{cccccccc}
&  & $^{10}$C & $^{12}$C & $^{14}$C & $^{16}$C & $^{18}$C & $^{20}$C    \\ 
\hline
cal.(ls=900 MeV) &$E_x(2^+_1)$ &
1.80 	&	1.45 	&	1.69 	&	0.40 	&	0.61 	&	0.57 	\\
&$E_x(2^+_2)$ &
2.60 	&	$-$	&	$-$	&	1.53 	&	0.83 	&	$-$	\\
\hline
cal.(ls=1500 MeV) &$E_x(2^+_1)$ &
1.95 	&	1.66 	&	3.75 	&	0.65 	&	0.87 	&	0.88 	\\
&$E_x(2^+_2)$ &
3.75 	&	$-$	&	$-$	&	2.82 	&	1.32 	&	$-$	\\
\hline
exp.&$E_x(2^+_1)$ &
3.354	&	4.439	&	7.012	&	1.766	&	1.62	&	$-$\\
&$E_x(2^+_2)$ &
6.58	&$-$&	$-$&	$-$&	$-$&	$-$\\
\end{tabular}
\end{center}
\end{table}

The results for the $E2$ transition strength are listed in Table.\ref{tab:e2}.
In the calculations, it is found that
$B(E2;2^+_1\rightarrow 0^+_1)$ drastically changes 
with the increase of neutron number in C isotopes. 
The theoretical $E2$ strength in $^{16}$C is very small, which  
is consistent with the recent measurement of the 
abnormally small $B(E2)$ in $^{16}$C \cite{Imai04}. 
The present results predict that,
also in $^{18}$C, $B(E2;2^+_1\rightarrow 0^+_1)$ is very small 
as the same order of that in $^{16}$C. It is interesting that 
the calculated $E2$ transition strength in $^{20}$C is 
not as small as those in $^{16}$C and $^{18}$C. 
The systematic change of the $B(E2;2^+_1\rightarrow 0^+_1)$ 
can be understood by the deformations of proton and neutron 
densities in the intrinsic states as described in the next section. 

\begin{table}
\caption{ \label{tab:e2} $E2$ transition strength in C isotopes
calculated by the VBP. The unit is
e$^2$fm$^4$. The experimental data for $^{16}$C is taken from 
Ref.\protect\cite{Imai04} and those for the other C 
are from Ref.\protect\cite{Raman}.}
\begin{center}
\begin{tabular}{cccccccc}
&  & $^{10}$C & $^{12}$C & $^{14}$C & $^{16}$C & $^{18}$C & $^{20}$C    \\ 
\hline
cal.(ls=900 MeV) &$B(E2;2^+_1\rightarrow 0^+_1)$ &
5.7 	&	7.2 	&	6.9 	&	1.9 	&	2.1 	&	5.3 	\\
&$B(E2;2^+_2\rightarrow 0^+_1)$ &
4.3 	&	$-$	&	$-$	&	4.7 	&	3.8 	&	$-$	\\
\hline
cal.(ls=1500 MeV) &$B(E2;2^+_1\rightarrow 0^+_1)$ &
5.4 	&	6.8 	&	5.9 	&	1.4 	&	0.6 	&	5.0 	\\
&$B(E2;2^+_2\rightarrow 0^+_1)$ &
3.4 	&	$-$	&	$-$	&	4.1 	&	4.9 	&	$-$	\\
\hline
exp.&$B(E2;2^+_1\rightarrow 0^+_1)$ &
12.4$\pm$ 0.2	&	8.2$\pm$ 0.1	&	3.74 $\pm$0.5	&	0.63	&$-$		&$-$		\\
\end{tabular}
\end{center}
\end{table}

\section{Discussion}\label{sec:discuss}

In this section, we analyze the intrinsic deformations 
of proton and neutron densities and discuss their effect 
on the observables such as the $E2$ transitions and radii.

\subsection{intrinsic deformation}

\begin{figure}
\noindent
\epsfxsize=0.49\textwidth
\centerline{\epsffile{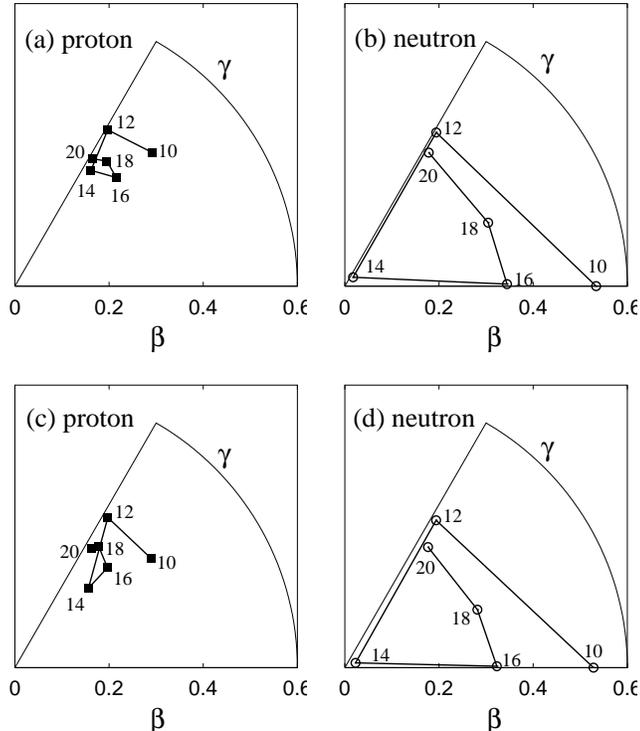}}
\caption{\label{fig:cdefo}
Deformation parameters $\beta,\gamma$ of the intrinsic states
calculated with $m=0.576$, $u_{ls}=900$ MeV (a,b) and 1500 MeV (c,d).
The mass numbers $A$ are written by the corresponding points.}
\end{figure}

We display in Fig.\ref{fig:cdefo} the deformation parameters $(\beta,\gamma)$
for the proton and neutron densities, which are defined by the moments
$\langle x^2 \rangle$, $\langle y^2 \rangle$, and 
$\langle z^2 \rangle$ of the intrinsic AMD wave function as,
\begin{eqnarray}
& \frac{\langle x^2 \rangle^{1/2}}{(\langle x^2 \rangle
\langle y^2 \rangle\langle z^2 \rangle)^{1/6}}\equiv 
\exp\left [ \sqrt\frac{5}{4\pi}\beta cos(\gamma+\frac{2\pi}{3})\right ],\\
& \frac{\langle y^2 \rangle^{1/2}}{(\langle x^2 \rangle
\langle y^2 \rangle\langle z^2 \rangle)^{1/6}}\equiv 
\exp\left [ \sqrt\frac{5}{4\pi}\beta cos(\gamma-\frac{2\pi}{3})\right ],\\
& \frac{\langle z^2 \rangle^{1/2}}{(\langle x^2 \rangle
\langle y^2 \rangle\langle z^2 \rangle)^{1/6}}\equiv 
\exp\left [ \sqrt\frac{5}{4\pi}\beta cos \gamma\right ].
\end{eqnarray}  
Here the $x$, $y$, and $z$ directions are chosen so as to satisfy
$\langle x^2 \rangle \le  \langle y^2 \rangle \le \langle z^2 \rangle$ and
$\langle xy \rangle =  \langle yz \rangle = \langle zx \rangle=0$.
As seen in Fig.\ref{fig:cdefo}, we find the drastic change of the neutron
deformation in C isotopes with the increase of the neutron number.
The neutron deformations are prolate, oblate and spherical in $^{10}$C,
$^{12}$C and $^{14}$C, respectively. In the neutron-rich region,
it become prolate again in $^{16}$C, and they are triaxial and oblate
in $^{18}$C and $^{20}$C, respectively. 
In contrast to the variation of neutron deformation, 
the proton deformations are rather stable. The deformation
parameters for the proton densities lie in the oblate region 
$\gamma\sim \frac{\pi}{3}$. These behavior are the same as the results of
Ref.\cite{ENYO-c10}.

\begin{figure}
\noindent
\epsfxsize=0.49\textwidth
\centerline{\epsffile{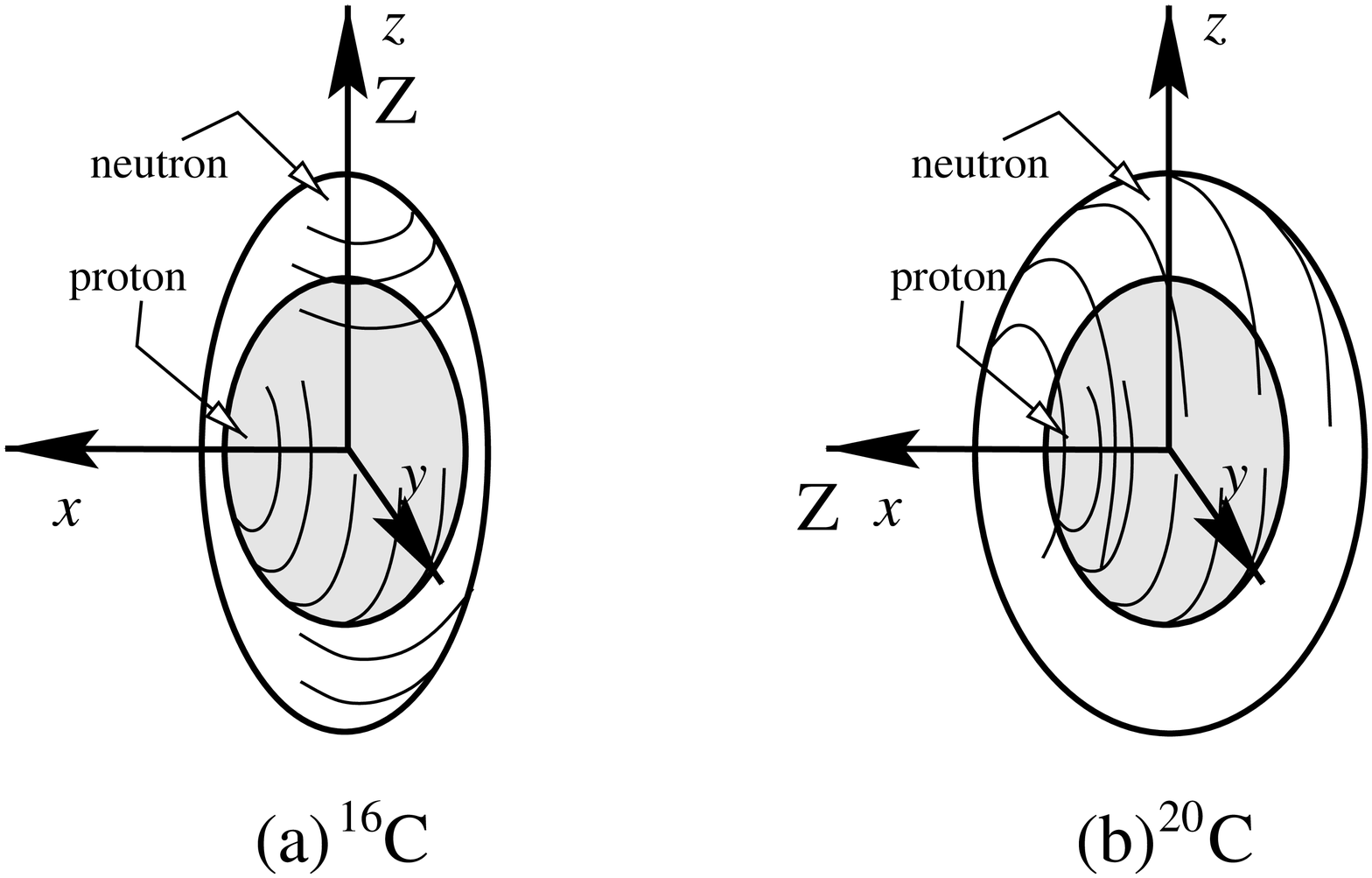}}
\caption{\label{fig:c16ob-pro}
Schematic figures for intrinsic deformations of 
the proton and neutron densities.
(a) the oblate proton and prolate neutron shapes in $^{16}$C, and (b) 
the oblate proton and oblate neutron densities in $^{20}$C. 
The $x$, $y$, and $z$ axes are chosen to be
$\langle x^2 \rangle \le \langle y^2 \rangle  \le \langle z^2 \rangle$.
The principal axis $Z$ for the ground band $J_Z=K^\pi=0$ is also displayed.}
\end{figure}

By analysing the component of the $K$-quantum states($P^J_{MK}\Phi^\pm_{AMD}$)
in the $2^+_1$, it is found that the $2^+_1$ state can be approximately
written by a single $K=0$ state when we chose a proper axis.
Then, we can define the (approximate) principal axis $Z$ in the body-fixed
frame and form the gound 
$K=0$ band with the $0^+_1$ and $2^+_1$ states in each C isotope.
In the systems, $^{12}$C, $^{14}$C and $^{20}$C, 
with the oblate or spherical neutron shapes, 
the principal axis $Z$ 
is the same as the symmetric axis $x$ which has the smallest
moment $\langle x^2 \rangle$ as shown in Fig.\ref{fig:c16ob-pro}(b). 
In other words, the dominant component of the 
excited state $2^+_1$ are the 
$J_Z=K=0$ state with respect to the symmetric axis $x$.
It is consistent with a naive expectation for the collective rotation.
It is notable that, in $^{10}$C, $^{16}$C and $^{18}$C, the 
deformations are different between proton and neutron densities.
In these nuclei, the symmetric axis for the proton shape differs from
that for neutron density. Namely, the symmetric axis of the oblate 
proton density is the $x$ direction, while that of the 
prolate neutron is the $z$ direction. The schematic figure of the proton and
neutron shapes in $^{16}$C is illustrated in Fig.\ref{fig:c16ob-pro}(a).
Such the configuration of the proton and neutron shapes is 
energetically favored because it has 
the maximum overlap between the proton and neutron densities.
Because of the coexistence of the different shapes between proton and neutron,
the second $2^+$ state appear due to the triaxiality of the 
total system.  
By analysing the component of the $K$-quantum states
($P^{J}_{MK}\Phi^\pm_{AMD}$) in the excited states, it is found that the 
$0^+_1$ and $2^+_1$ states belong to the ground $K^\pi=0^+$ band 
and the $2^+_2$ state is classified into the side-band $K^\pi=2^+$
when we regard the $z$ direction with the largest moment 
$\langle z^2\rangle$ as the principal axis $Z$
as noted in Fig.\ref{fig:c16ob-pro}(a).
It is important that the principal axis $Z$ is not 
the same as the symmetric axis $x$ for the proton density 
but is perpendicular to
the $x$ in these nuclei.
The existence of the side band $K^\pi=2^+$ 
in $^{16}$C and $^{18}$C has not been experimentally confirmed yet.
Concerning $^{10}$C, the triaxiality of the mirror nucleus $^{10}$Be
were discussed in Ref.\cite{ITAGAKI-tri},
and the known $2^+_2$ state in $^{10}$Be is assigned 
to be the band-head state of the side band $K^\pi=2^+$ 
\cite{ITAGAKI-tri,ENYOg}.

We compare the matter deformation of the present results 
with a HF+BCS calculation with Skyrme force by Tajima
et al.\cite{Tajima96}. In the HF+BCS calculation, the C isotopes 
in the $A \le 14$ region have spherical shapes, which contradict to 
the knowledge that $^{12}$C is oblately deformed.
It is natural because the mean-field calculation is considered not to be 
valid for very light nuclei.
In the neutron-rich C, the calculated 
quadrupole deformation parameter 
is positive in $^{16}$C and $^{18}$C and oblate in $^{20}$C
according to the HF+BCF calculation.
The general behavior of the quadrupole deformation of the matter density 
in neutron-rich C
seems to be similar with the present results, though the different shapes
between proton and neutron hardly appear in the HF calculations.

\subsection{$E2$ transition}

The intrinsic deformation is closely related to the $E2$ transition strength.
Since the present results indicate that 
the proton radius does not drastically change with the increase 
of the neutron number in C isotopes as shown in Fig.\ref{fig:crmsr}, 
the $B(E2)$ is dominantly 
determined by the deformations.

As mentioned before, the $0^+_1$ and $2^+_1$ states belong to the 
ground $K^\pi=0^+$ band, where we regard the $x$ axis as the principal 
axis $Z$ in $^{12}$C, $^{14}$C and $^{20}$C, and the $z$ axis as the
$Z$ axis in $^{10}$C, $^{16}$C and $^{18}$C.
In order to link the intrinsic deformations with the $B(E2)$,
we remind the reader the well-known approximate relation between the 
$B(E2)$ and the intrinsic quadrupole moment $Q_0$:
\begin{equation}
B(E2;2^+_1\rightarrow 0^+_1)=\frac{1}{16\pi} e^2 Q_0^2.
\end{equation}
The intrinsic quadrupole moment $Q_0$ here is defined with respect to 
the principal axis as 
$Q_0=2\langle Z^2\rangle -\langle X^2\rangle -\langle Y^2\rangle$
, and is related to the deformation parameter $\beta_p$, $\gamma_p$
for the proton density.
In $^{12}$C, $^{14}$C and $^{20}$C, where $Z\approx x$ and $\gamma\approx 
\pi/3$, $Q_0$ is approximated as:
\begin{equation}
Q_0=2\langle x^2\rangle -\langle y^2\rangle -\langle z^2\rangle
\approx -\sqrt{\frac{5}{4\pi}}N_pe\beta_p r_e^2,
\end{equation}
where $N_p$ and $r_e$ are the proton number and 
the root-mean-square charge radius.

In $^{10}$C, $^{16}$C and $^{18}$C with $Z\approx z$,
$Q_0$ depends on the deformation parameter $\gamma_p$ as:
\begin{equation}\label{eq:c16q0}
Q_0=2\langle z^2\rangle -\langle x^2\rangle -\langle y^2\rangle
\approx \sqrt{\frac{5}{4\pi}}N_p e\beta_p \cos \gamma_p r_e^2.
\end{equation}
The important point is that the effect of the proton deformation 
on the $Q_0$ decreases in these nuclei
because of the factor $\cos \gamma_p$. 
Especially, in $^{16}$C and $^{18}$C, the deformation parameter 
$\gamma_p\sim \pi/3$ makes
the $Q_0$ very small. This is the reason for the unusually small
$B(E2;2^+_1\rightarrow 0^+_1)$ in $^{16}$C and $^{18}$C 
shown in Table.\ref{tab:e2}.
In other words, the $B(E2;2^+_1\rightarrow 0^+_1)$ in $^{16}$C and $^{18}$C 
is reduced due to the deviation the principal axis $Z$ of the total system
from the symmetric axis $x$ for 
the proton density.
The origin of the deviation is the different shapes between proton and neutron
densities. 
On the other hand, it is interesting that 
the larger $B(E2:2^+_1\rightarrow 0^+_1)$ is predicted in $^{20}$C,
because it has the oblate proton and neutron shapes, therefore,  
the principal axis aligns to the symmetric axis $x$ 
as shown in Fig.\ref{fig:c16ob-pro}(b).

The hindrance of the $B(E2:2^+_1\rightarrow 0^+_1)$ in $^{16}$C 
can be described also from the point of view of the collective rotation.
The $2^+_1$ state of $^{16}$C is roughly regarded to be the 
state with $J_z=K=0$. The $J=2,J_z=0$ state is given by 
the linear combination of 
the $J_x=2$ state and the $J_y=2$ state, which are rotating around the 
$x$ and $y$ axes, respectively. The rotation around $x$ axis
causes no proton excitation, therefore, gives no contribution 
on the $B(E2)$, because $x$ is the symmetric axis of proton density.
As a result, the $J_x=2$ component reduces 
the $B(E2:2^+_1\rightarrow 0^+_1)$. In contrast to the small
$B(E2:2^+_1\rightarrow 0^+_1)$, it is predicted that 
the $B(E2:2^+_2\rightarrow 0^+_1)$ in the side band is large
because the $2^+_2$ state is dominated by the $J_z=2$ state,
where the oblate shape of the proton densities gives proton excitations.

As mentioned above, the abnormal small 
$B(E2:2^+_1\rightarrow 0^+_1)$ in $^{16}$C can be understood by the
difference between oblate proton and prolate neutron shapes. 
Even if the proton shape in $^{16}$C is spherical 
or slightly prolate, the small $B(E2:2^+_1\rightarrow 0^+_1)$ can be 
also described. The characteristic of the present result is the stable
proton structure in the series of C isotopes and the prediction of the
second $2^+$ state with large $B(E2)$ in the side band.
In order to conclude the intrinsic shapes of proton and neutron densities,
we need further experimental information such as the systematics of $B(E2)$ 
in other neutron-rich nuclei and some probes for the side-band in $^{16}$C.

\subsection{radii}
In Fig.\ref{fig:crmsr}, we present the calculated results of
matter, proton and neutron radii of C isotopes 
while comparing them with the experimental radii which are derived from the 
interaction cross sections\cite{Ozawa}.
As seen in Fig.\ref{fig:crmsr}(b), the proton radius does not 
drastically change with the increase of neutron number, 
while the neutron radius increases rapidly in the neutron-rich 
$A\ge 16$ region. 
The gap of the matter radii between $^{14}$C and $^{16}$C is found to 
originate from a gap in neutron radii. The reason for the gap is described by
the neutron deformations as follows. The neutron density 
is compact in $^{14}$C
because it has a spherical shape due to the neutron shell closure.
On the other hand, in $^{16}$C, the neutron radii is large because 
of the prolate deformation.

The calculated matter radii systematically reproduce the 
experimental data. The large matter radii in the neutron-rich region 
originate from the enhancement of the neutron radii.
In contrast to the variation of the neutron radii, 
the proton radii is stable and is generally compact.
As a result of the stable proton structure, 
the neutron skin structure enhances in the neutron-rich C
isotopes. 
The development of neutron-skin in the C isotopes found in the present results
is consistent with the results of mean-field calculations \cite{Tajima96}.
In the calculations by Thiamova et al.
\cite{Thiamova03} with AMD+GCM,
the radii of neutron-rich C are somehow underestimated. This is 
considered to be because of the lack of the three-body term in the 
effective force in their calculations. 

\section{Results of VAP calculation}\label{sec:vap}

So far, we discuss the structure of C isotopes based on the VBP calculations.
As mentioned before, the quantitative reproduction of the excitation energy 
$E_x(2^+_1)$ in the VBP calculations is not satisfactory in the neutron-rich 
C. Instead, the VAP calculation\cite{ENYOe,ENYOg} is more useful to 
describe the detail of the level spacing. 
We perform energy variation after the spin-parity projection for 
the $0^+_1$, $2^+_1$ and $2^+_2$ states of $^{16}$C, and obtain 
three independent AMD wave functions. We evaluate the observables by
diagonalizing the Hamiltonian and norm matrices with respect to the 
three wave functions as done in Refs.\cite{ENYOe,ENYOg}.

\begin{figure}
\noindent
\epsfxsize=0.49\textwidth
\centerline{\epsffile{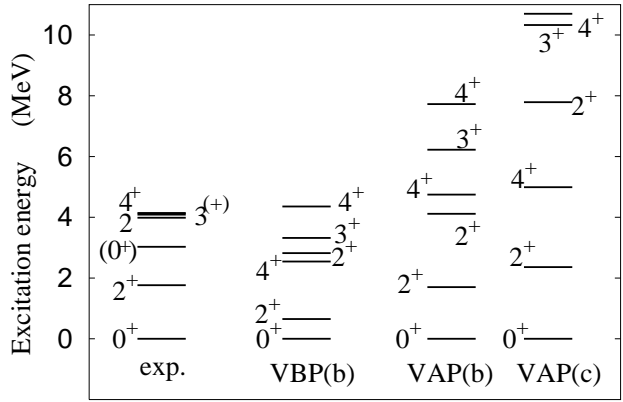}}
\caption{\label{fig:c16ene}
Level scheme of the low-lying states of $^{16}$C.
The theoretical results obtained by the 
VAP and VBP calculations with the interaction parameters
(b)case 3 of MV1 with $m=0.576$, $u_{ls}=1500$ MeV and (c)case
1 of MV1 with $m=0.62$, $u_{ls}=3000$ MeV
are illustrated with the experimental data.
}
\end{figure}

Figure \ref{fig:c16ene} shows the level scheme of the low-lying states of 
$^{16}$C obtained by the VBP and VAP calculations with 
the interaction parameters (b) case 3 of MV1 with $m=0.576$, $u_{ls}=1500$ MeV.
We also show the VAP results obtained by the other parameter set (c) 
case 1 of MV1 with $m=0.62$, $u_{ls}=3000$ MeV, 
which were adopted in Ref.\cite{ENYOe}
to reproduce well the level structure of $^{12}$C. 
The level spacing between $0^+_1$ and $2^+_1$ is well reproduced 
by the VAP calculations. The excitation energies of the 
side-band $K^\pi=2^+$ states rise comparing those with VBP. 
Especially, in the VAP with the interaction (c), the $2^+_2$ state becomes
relatively high because of the strong spin-orbit force.

The results of the $E2$ transition strength in $^{16}$C are listed in 
Table.\ref{tab:c16be2}. Comparing the theoretical values with the 
experimental data, the VAP calculations tend to overestimate 
$B(E2;2^+_1\rightarrow 0^+_1)$. 
It should be noted that the triaxial shape $\gamma_p\sim 6/\pi$ 
of the proton density is found in the VAP results.
This is the reason for the larger 
$B(E2;2^+_1\rightarrow 0^+_1)$ in VAP than that in VBP in which
$^{16}$C has an oblate proton shape $\gamma_p\sim 3/\pi$. Because of the
triaxial proton shape in VAP,
the $E2$ transition in the side band, $B(E2;2^+_2\rightarrow 0^+_1)$,
is smaller than that of VBP.

In the present VAP calculations of $^{16}$C, 
the $0^+_1$ and $2^+_1$ level spacing
is well reproduced,
while the reproduction of the small $B(E2;2^+_1\rightarrow 0^+_1)$ is not 
satisfactory.
In order to insight the structure of $^{16}$C, we need 
to improve wave functions in more detail. Moreover,
the further experimental information for other excited states
is helpful to determine the 
proton shape in $^{16}$C.

\begin{table}
\caption{\label{tab:c16be2} The $E2$ transition strength
in $^{16}$C. The theoretical results are obtained by the
VAP and VBP calculations with the interaction parameters
(b) $m=0.576$, $u_{ls}=1500$MeV and (c) $m=0.62$, $u_{ls}=3000$MeV.
The unit is e$^2$fm$^4$.}
\begin{center}
\begin{tabular}{ccccc}
& exp. & VBP(b) & VAP(b) & VAP(c) \\ 
$B(E2;2^+_1\rightarrow 0^+_1)$ & 0.63 & 1.4  &  3.7  & 2.7 \\
$B(E2;2^+_r\rightarrow 0^+_1)$ & $-$  & 4.1  &  2.0  & 2.6 \\
\end{tabular}
\end{center}
\end{table}

\section{Summary}\label{sec:summary}

We studied the structure of even-even C isotopes
with the AMD method.
The experimental data of the 
binding energies, $B(E2)$, and radii of C isotopes are well reproduced 
in the present calculations. 
Systematic analysis of the proton and neutron shapes
in C isotopes was done based on the VBP calculations of AMD.
The results indicate that the neutron shape drastically changes with the 
increase of neutron number, while the proton shape is rather stable.
It is suggested that the difference between proton and neutron shapes may
appear in $^{16}$C and $^{18}$C as well as $^{10}$C. 
The $E2$ transition strength, $B(E2)$, was discussed 
in relation to the deformation.
The unusually small $B(E2;2^+_1\rightarrow 0^+_1)$ in $^{16}$C, which has been
recently measured, was described by the coexistence of the 
oblate proton shape and the 
prolate neutron shape. According to the present prediction, 
the $B(E2;2^+_1\rightarrow 0^+_1)$ in $^{18}$C is as small as that 
in $^{16}$C, while the $B(E2)$ is larger in $^{20}$C.
The deviation between the proton and neutron shapes 
play an important role in the small $B(E2)$. 
The present results show the enhancement of the neutron skin structure 
in neutron-rich C.
It was found that the stable proton structure in C isotopes plays an 
important role in the neutron skin structure as well as
in the systematics of $B(E2)$. 

In order to extract a naive picture on the proton and neutron shapes
we applied the simplest version of AMD, based on a single AMD wave function. 
The quantitative reproduction of both the excitation energy($E_x(2^+_1)$)
and $B(E2)$ was not satisfactory in the present work.
We consider that more detailed investigations with the 
improved wave functions as well as with the use of appropriate 
effective nuclear forces are required.

In the experiment of the $^{208}$Pb+$^{16}$C inelastic scattering, 
the contributions of nuclear excitation and Coulomb excitation 
from the ground state to the $2^+_1$(1.77 MeV) state were
analyzed \cite{Elekes04}. 
The ratio of the neutron and proton transition matrix elements 
$M_n/M_p$ implies the neutron excitation is dominant in the 
$2^+_1$ state in $^{16}$C.
This is consistent with the 
present results of oblate proton and prolate neutron shapes.
Unfortunately, with this experimental information for 
the excitation to the $2^+_1$(1.77 MeV), 
it is difficult to know whether the proton deformation is 
oblate or slightly prolate (or slightly triaxial), because in both cases 
the sign of the $M_p(0^+_1\rightarrow 2^+_1)$ is same. Namely, even if the
proton density has an oblate shape, the $M_p$ is not negative but a positive
value because of the different orientations of the symmetric axes 
between the proton and neutron shapes as shown in $Q_0$ given by 
Eq.\ref{eq:c16q0}.
The characteristic of the present result is the stable
proton structure and systematics of $B(E2)$ in the series of C isotopes. 
If the $^{16}$C has the oblate proton shape, 
the second $2^+$ state appears.
We should stress that, whether the proton density is 
oblate or slightly prolate in $^{16}$C, the proton shape must differ from the 
large prolate deformation of the neutron to decribe the small $B(E2)$.
It is concluded that the small $B(E2)$ indicates 
the difference between proton and neutron shapes in $^{16}$C.   
In order to understand the details of the 
intrinsic shapes of proton and neutron densities,
we need more systematic analysis of C isotopes 
with the help of the further experimental information 
such as the $B(E2)$ in other neutron-rich nuclei, $^{18}$C and $^{20}$C,
and information for the side-band in $^{16}$C.

\acknowledgments

The author would like to thank Prof. H. Horiuchi, Dr. M. Takashina,
and Dr. N. Itagaki for many discussions. She is  also thankful to 
Prof. T. Motobayashi, Dr. N. Imai and Z.Elekes for valuable comments. 
The computational calculations in this work were supported by the 
Supercomputer Project Nos. 58 and 70
of High Energy Accelerator Research Organization(KEK).
This work was supported by Japan Society for the Promotion of 
Science and a Grant-in-Aid for Scientific Research of the Japan
Ministry of Education, Science and Culture.
The work was partially performed in the ``Research Project for Study of
Unstable Nuclei from Nuclear Cluster Aspects'' sponsored by
Institute of Physical and Chemical Research (RIKEN).

\end{document}